\documentclass{aa}

\usepackage{graphicx}
\usepackage[varg]{txfonts}
\usepackage{natbib}
\usepackage{hyperref}

\begin{document}

 \title{Investigation of nulling and subpulse drifting properties of PSR J1727$-$2739}

 \author{Z. G. Wen\inst{1,2},
         N. Wang\inst{1,3},
         J. P. Yuan\inst{1,3},
         W. M. Yan\inst{1,3},
         R. N. Manchester\inst{4},\\ 
	 R. Yuen\inst{1,3},
	 V. Gajjar\inst{1,3}}

   \institute{Xinjiang Astronomical Observatory, 150, Science-1 Street, Urumqi, Xinjiang, 830011, China \\
              \email{na.wang@xao.ac.cn} \and
              University of Chinese Academy of Sciences, 19A Yuquan road, Beijing, 100049, China \and
              Key laboratory of Radio Astronomy, CAS, Nanjing, 210008, China \and
	      CSIRO Astronomy and Space Science, PO Box 76, Epping NSW 1710, Australia}


\abstract{}{}{}{}{}

  \abstract
   {}
   {To make a detailed study of the
   nulling and subpulse drifting in PSR J1727$-$2739 for investigation
   of its radiation properties.  }
   {The observations were carried out on
   20 March, 2004 using the Parkes 64-m radio telescope, with a
   central frequency of 1518 \rm{MHz}. A total of 5568 single pulses
   were analysed.}
   {This pulsar shows well defined nulls
   with lengths lasting from 6 to 281 pulses and separated by
   burst phases ranging from 2 to 133 pulses. We estimate a
   nulling fraction of around 68\%. No emission in the average pulse
   profile integrated over all null pulses is detected with
   significance above 3$\sigma$. Most transitions from nulls to bursts
   are within a few pulses, whereas the transitions from bursts
   to nulls exhibit two patterns of decay: decrease gradually or
   rapidly. In the burst phase, we find that there are two
     distinct subpulse drift modes with vertical spacing between the drift
     bands of $9.7 \pm 1.6$ and $5.2 \pm 0.9$ pulse periods,
     while sometimes there is a third mode with no subpulse
     drifting. Some mode transitions occur within a single burst,
     while others are separated by nulls.  Different modes have
     different average pulse profiles. Possible physical mechanisms
     are discussed.}
   {}
   \keywords{Stars: neutron --
	     Pulsar: individual: PSR J1727$-$2739
             }
   \authorrunning{Z. G. Wen et al.}
   \titlerunning{Nulling and Subpulse Drifting of PSR J1727$-$2739}

   \maketitle
%

\section{Introduction}
\label{intro}
\begin{figure}
\begin{center}
\includegraphics[width=80mm,height=100mm]{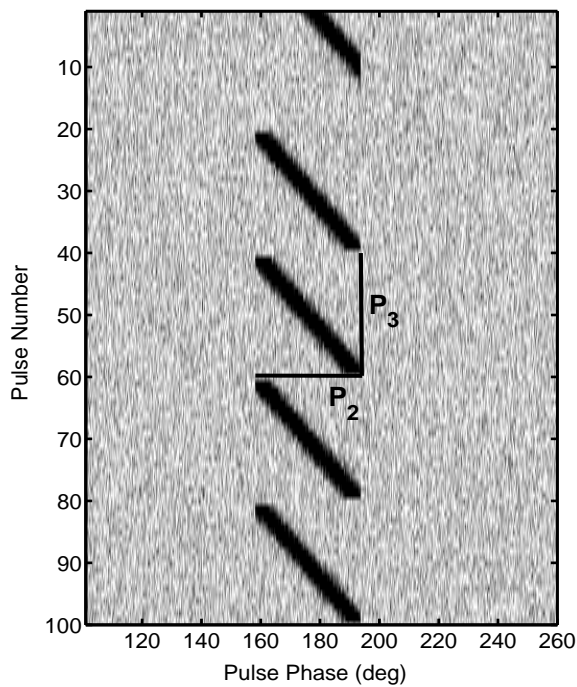}
\caption{The schematic diagram for a stack of one hundred
  successive pulses of fake data, with pulse phase plotted horizontally
  and pulse number vertically. $P_2$ and $P_3$ are the separation of
  successive drift bands horizontally and vertically, respectively.}
\label{fake}
\end{center}
\end{figure}

It is well known that pulsar emission shows extremely complicated
properties such as fluctuations in the pulse width, intensity and
phase. Systematic drifting of subpulses across the emission window was
discovered in 1968 \citep{dra68}. \citet{wel06} reported that at least
one third of pulsars possess drifting subpulses, indicating that this
phenomenon is a common behaviour in pulsars. As shown in Fig.~\ref{fake}, three
parameters for describing the observed drift of subpulses in a
phase-time diagram are the horizontal time interval between successive drift bands
($P_2$), the vertical band spacing at the same pulse phase ($P_3$), and the
drifting rate ($\Delta\phi = P_2 / P_3$). Some pulsars show several
drift modes with different $P_3$. For instance, PSR B0031$-$07
exhibits three distinct and stable drift modes at low frequency
\citep{hu70}. The transition between modes can be rapid within one or
a few pulse periods \citep{smi05}.

Nulling is when the pulsed emission abruptly switches off with duration ranging from one
to many thousands of pulse periods and then abruptly resumes its
normal flux density \citep{bac70}, and is thought to be magnetospheric
in origin \citep{kra06,wan07,tim10}. The phenomenon seems to occur at
all frequencies and to all pulse components simultaneously, 
but partial nulls are also detected \citep{wan07}.
A variety of
models have been proposed for nulling, such as orbital companions
\citep{cor08}, missing the line of sight \citep{her07}, switching between
curvature radiation and inverse Compton scattering \citep{zha97} and
magnetic field instability \citep{gep03}.

Diverse patterns of transition between burst and null states have
been observed in several pulsars. For PSR B0818$-$41 \citep{bha10} and
PSR J1502$-$5653 \citep{li12}, an abrupt rise of intensity when
emission starts after a null and an exponential decay of pulse
emission at the end of a burst are seen. On the
other hand, an abrupt onset of nulling was observed from PSRs
B0031$-$07 and J1738$-$2330 \citep{viv95,vis14}.

Studies of pulsars that exhibit both nulling and drifting subpulses
and the interaction between them provide insights into the 
properties of pulsar magnetospheres \citep{red05,for10}. Attempts have been made to unite
the two phenomena \citep{van02,van03,jon11} by examining more pulsars
with similar behaviour. \citet{smi05} reported that transitions from
one drift mode to another are interspersed with at least one null
pulse. Different drift modes were also detected from PSR B0809+74,
which exhibits transitions between two distinct modes after it goes
through a null \citep{van02}. PSR B1737+13 provides a unique
opportunity to study the interactions between drifting subpulses and
nulling behaviours with its two intensity modulation periods of 
9 and 92 rotation periods, which suggest a subbeam 
carousel with 10 'beamlets' \citep{for10}. Investigations of the on and off states for PSR
B1931+24 which are accompanied by variations in spin-down rate show
that nulling is a global property in pulsar
magnetospheres \citep{kra06}. Furthermore, studies \citep{lyn83,van02,jan04,red05}
also revealed that the occurrence of nulls is not random and may be
governed by emission cycles. 
\citet{wan07} showed
the correlation of pulsar nulling and mode changing and proposed that
those two emission properties may result from large-scale and
persistent changes in the magnetospheric current
re-distribution. \citet{ran03} pointed out that pulsar magnetosphere
is a non-steady, and non-linear interactive system. So far nulling has
been detected in almost 200 pulsars and the duration of a null and the
interval between them appear random, although a correlation between
the frequency of occurrence and the pulsar rotation period has
been suggested \citep{rit76}. Despite these studies, however, a
comprehensive understanding is still lacking and it is still unclear
to what extent the two phenomena are related.

In this paper, we focus on PSR J1727$-$2739, which was discovered in
the Parkes Multibeam Pulsar Survey \citep{hob04}, with measured
rotational period of $P=1.29\ \rm{s}$ and first period derivative 
$\dot P=1.1 \times 10^{-15}\ \rm{s\ s^{-1}}$, giving the
characteristic age of $1.86 \times 10^7\ \rm{yr}$ and the surface
magnetic field of $10^{12}\ \rm{G}$, respectively \citep{hob04}. 
In Section \ref{sec2}, we describe the
observations and detailed analysis of nulling and drifting subpulses
for this pulsar. Examinations of the nulling and the properties of the
bright states before and after the nulls, together with different
modes of subpulse drift rates are also discussed. In section \ref{sec3},
we discuss several possible mechanisms for the phenomena and summarise our results.


\section{Analysis and Results}
\label{sec2}
\begin{figure}
\begin{center}
\includegraphics[width=80mm]{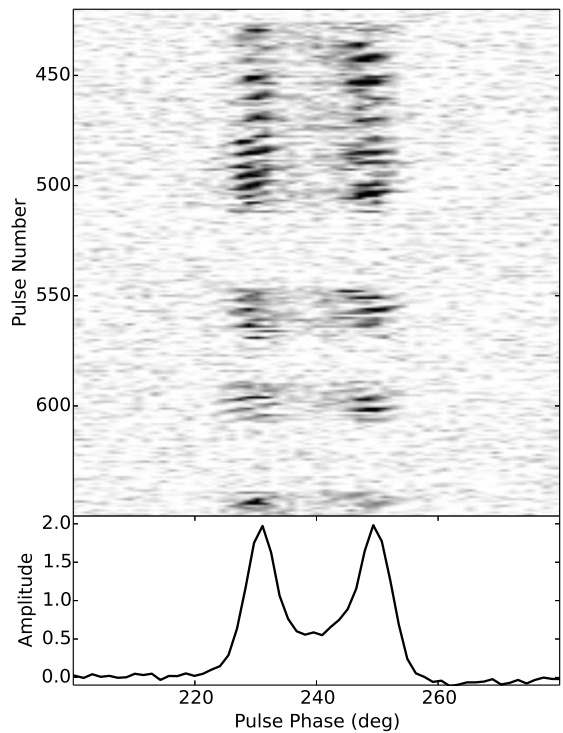}
\caption{A sample of single pulse sequence in grey-scale intensities
  for PSR J1727$-$2739, in which increasing darkness represents higher
  intensity. Frequent nulling and subpulse drifting are observed. The
  pulse profile averaged over the whole data span is shown in the lower panel.}
\label{grey_plot}
\end{center}
\end{figure}

The observations were made on 20 March, 2004 using the Parkes 64-m
radio telescope in a frequency band centered at 1518 \rm{MHz}. The 
data used for investigation of nulling behaviour are
sampled every 250 \rm{$\mu$s} for 2 hours and contain 5568 pulse
periods. A more detailed description of the observing system is given
by \citet{wan07}. In order to increase signal-to-noise ratio, 
(S/N), the time series is averaged to 256 phase bins per
pulse period. The average pulse profile, which is obtained by integrating 
across the data span, is illustrated in the lower panel of
Fig. \ref{grey_plot}. Two steep-sided components with approximately
equal amplitude are separated by $32\degr.0 \pm 0\degr.7$ and are
joined by a saddle region of emission.

The upper panel of Fig. \ref{grey_plot} shows a sample of single pulse
data demonstrating both frequent nulling and complicated drifting
subpulse phenomena. Instances of null states can easily be identified
between pulse number 514 to 545, pulse number 570 to 587 and pulse
number 608 to 638. For the burst state, strong pulses are clustered,
between which short null intervals are seen. Analysis of the
single-pulse sequences also reveals regions of drifting patterns with
the variable drift rates. The drift rate at the leading region is
steeper than that at the trailing region. On several occasions, the
drifting at the trailing component becomes irregular, and for the
first and last few active pulses the drift behaviour disappears,
which indicates that it takes a few pulse periods for the
  drifting to establish and then cease before a null.

\subsection{Nulling}

\begin{figure}
\begin{center}
\includegraphics[width=95mm]{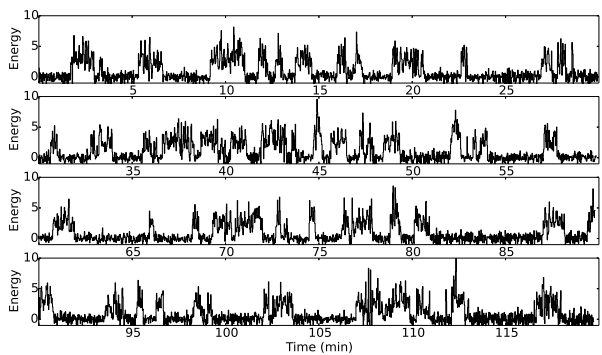}
\caption{The two-hour evolution of relative pulse energy for PSR
  J1727$-$2739. The time sequence is equally divided into four panels,
  each presents half an hour of data.}
\label{time_line}
\end{center}
\end{figure}

\begin{figure}
\begin{center}
\includegraphics[width=95mm]{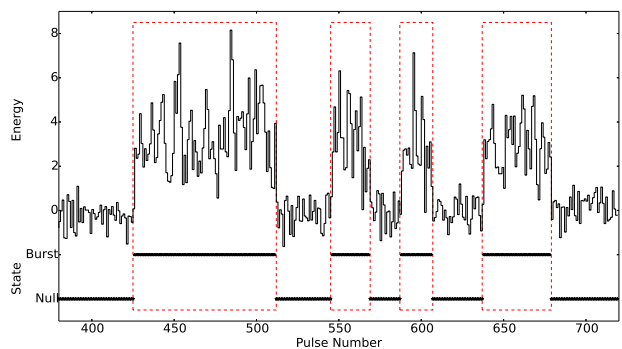}
\caption{Upper part: On-pulse energy versus pulse number. For clarity, four identified burst states are
  enclosed in rectangular regions. The null or burst states are
  classified and shown in the lower part of figure.}
\label{tsline}
\end{center}
\end{figure}

Fig. \ref{time_line} shows the time sequence of pulse energy
  for PSR J1727$-$2739 revealing many blocks of consecutive strong
  pulses separated by frequent nulling. The individual
  on-pulse energy was calculated by averaging the baseline-subtracted
  data in the on-pulse window and then scaling by the average
  value. The average off-pulse energy was measured in an equal number
  of off-pulse bins and then scaled by the same factor.  We follow
the similar procedure presented by \citet{bha10} for identification of
individual nulls, in which pulses with intensity smaller than $5
\sigma_{ep}$ are classified as nulls, where $\sigma_{ep}$ is the
uncertainty in the pulse energy calculated from the
  root-mean-square energy in the off-pulse window. Fig. \ref{tsline}
shows the on/off pulse time series for a section of the data with
individual pulses classified as either null or burst states. In this
work, a total of 53 blocks of burst state (2085 pulses) are identified
for the whole observations. The durations of burst states vary from
about 2 pulses to 133 pulses with a median duration of 28 pulses,
although occasional individual strong pulses are also seen within
(otherwise) null intervals.  Null states last from about 6 pulses to
281 pulses with a median duration of 43 pulses.

\begin{figure}
\begin{center}
\includegraphics[width=90mm,height=80mm]{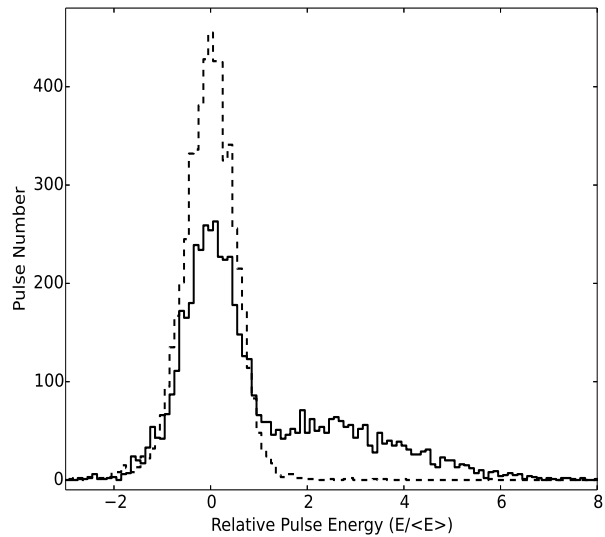}
\caption{Energy histograms for the on-pulse windows (solid
    line) and off-pulse windows (dashed line). The energies are
    normalized by the mean on-pulse energy. }
\label{energy}
\end{center}
\end{figure}

The cryogenically-cooled receiver had a stable system
  temperature over two hours and the variation of the system gain is
  small, so we use uncalibrated data for our fluctuation analyses
  \citep{wan07}. Pulse energy distributions provide statistical
information for the characterisation of pulsar nulling
properties. Fig. \ref{energy} presents the energy distributions for
the on-pulse windows and off-pulse windows. The off-pulse energy
histogram shows a Gaussian shape centered around zero, while the
broader on-pulse energy histogram has two Gaussian components
corresponding to active and null pulses. These two histograms clearly
overlap. Following the method described by \citet{rit76}, an
increasing fraction of the off-pulse histogram centered on zero energy
was subtracted from the observed on-pulse distribution until the sum
of the difference counts in bins with $E < 0$ was zero. The fraction
determined in this way is the ``null fraction'' (NF) which is the
fraction of the pulse sequence which is null. The uncertainty of NF is
simply given by $\sqrt{n_p} / N$, where $n_p$ is the number of null
pulses and N is the total number of observed pulses \citep{wan07}. The
null fraction (NF) of this pulsar is estimated to be $68.2 \pm 1.1
\%$.

\begin{figure*}
\begin{center}
\includegraphics[width=80mm]{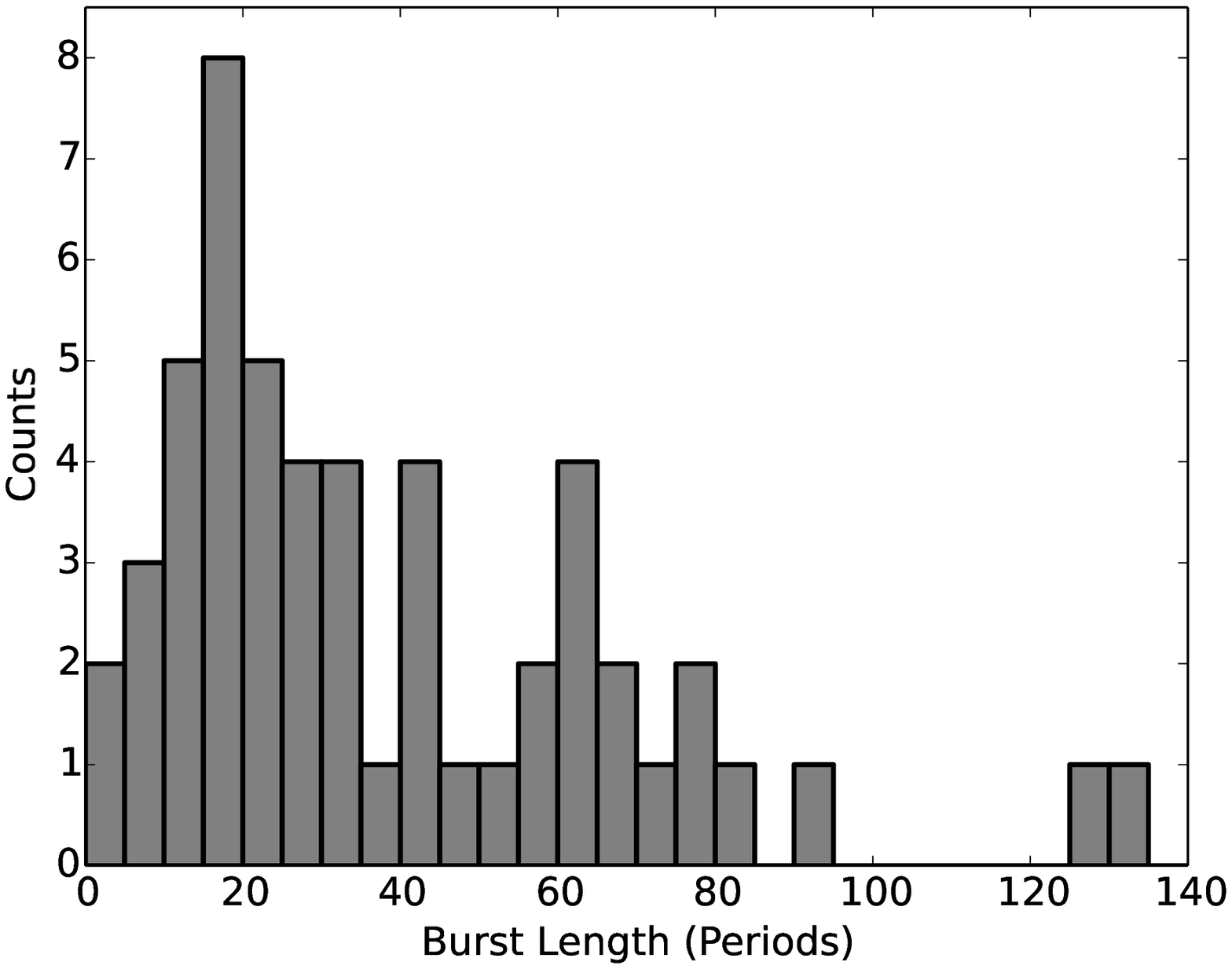}
\hspace{3ex}
\includegraphics[width=80mm]{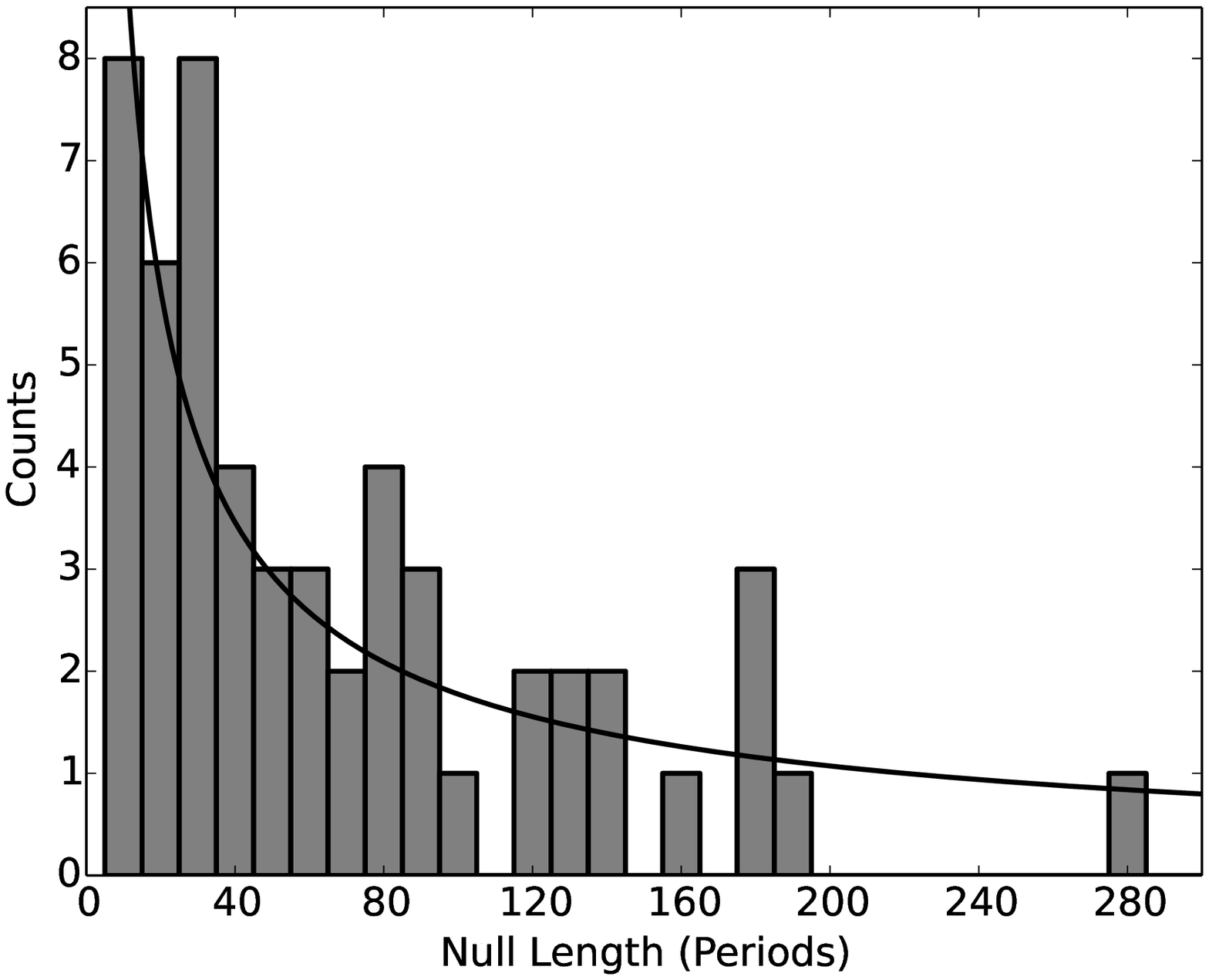}
\caption{The distributions of duration of burst (left panel) and null
  (right panel) states. The histogram of null duration has an approximate
  power-law distribution with a slope of $\alpha = -1.0 \pm 0.2$,
  as indicated with the solid curve in the right panel.}
\label{duration}
\end{center}
\end{figure*}

The identified contiguous null and burst lengths are shown in
Fig. \ref{duration}. It is obvious that the histogram of burst length peaks at around 20 pulse periods. For the short nulls, the histogram indicates that the distribution has a favor of around 5$-$35 periods.
The decline of short nulls to long nulls may be
fitted by a power-law distribution which has a slope of $\alpha = -1.0
\pm 0.2$.

\subsection{Pulse energy variations}

\begin{figure}
\begin{center}
\includegraphics[width=80mm]{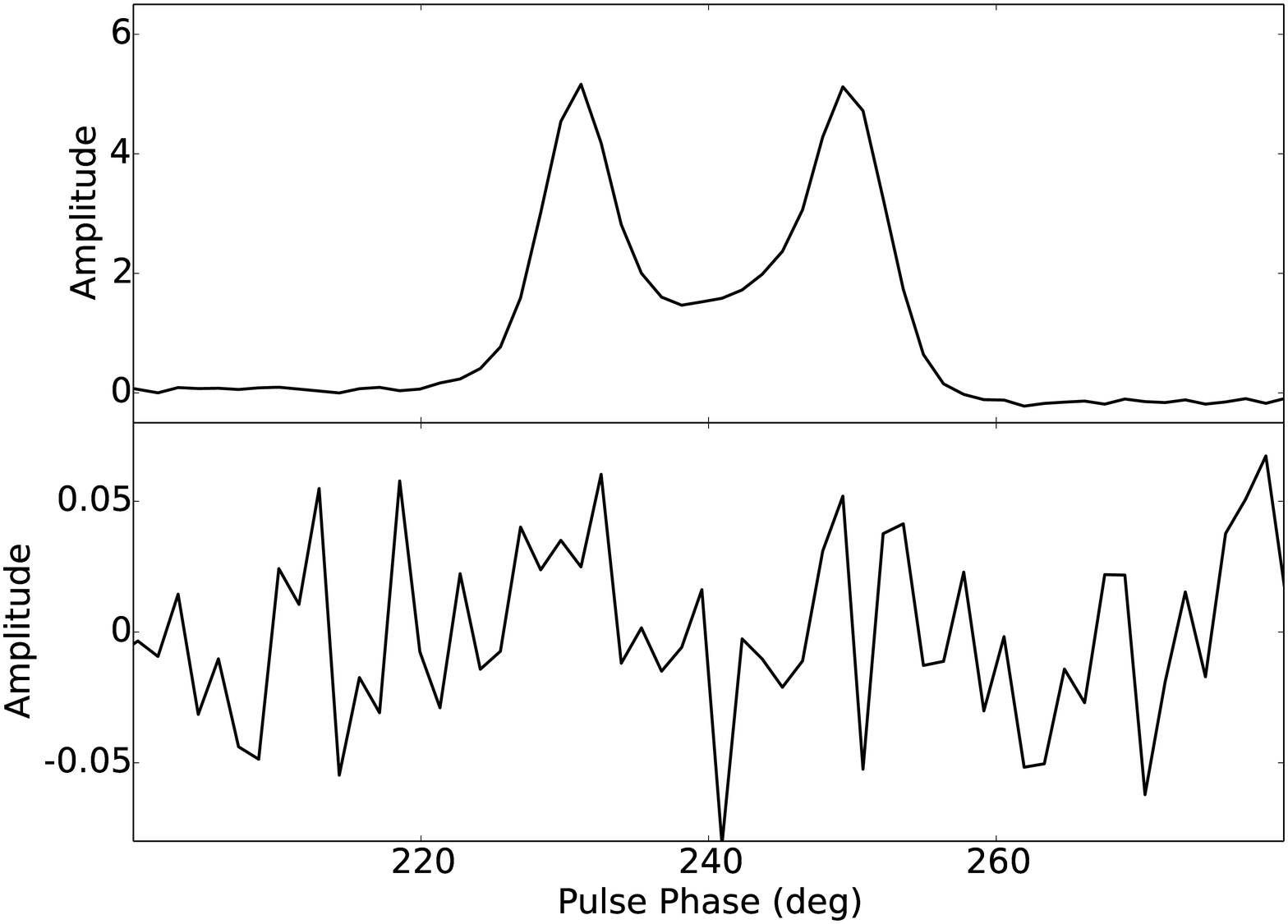}
\caption{Integrated pulse profiles of PSR J1727$-$2739 for 2119 burst
  pulses (top panel) and the 3449 null pulses (bottom panel).}
\label{profiles}
\end{center}
\end{figure}

Through a careful inspection of the null sequences, three examples of
an isolated detectable pulse were found during null states, which
means that, technically, "bursts" can be as short as 1 pulse period.
Also, detectable pulses were found in nominally null regions adjacent
to burst regions. A "guard band" of 1 pulse before and after burst
regions was used to eliminate 31 of these transition pulses. After
removing these 34 pulses, there is no obvious energy with
S/N larger than 3 after integrating all 3449
null pulses, as shown in the lower panel of Fig. \ref{profiles}. The
integrated pulse profile obtained from all 2119 burst pulses is shown
in the upper panel.

\begin{figure}
\begin{center} 
\includegraphics[width=95mm]{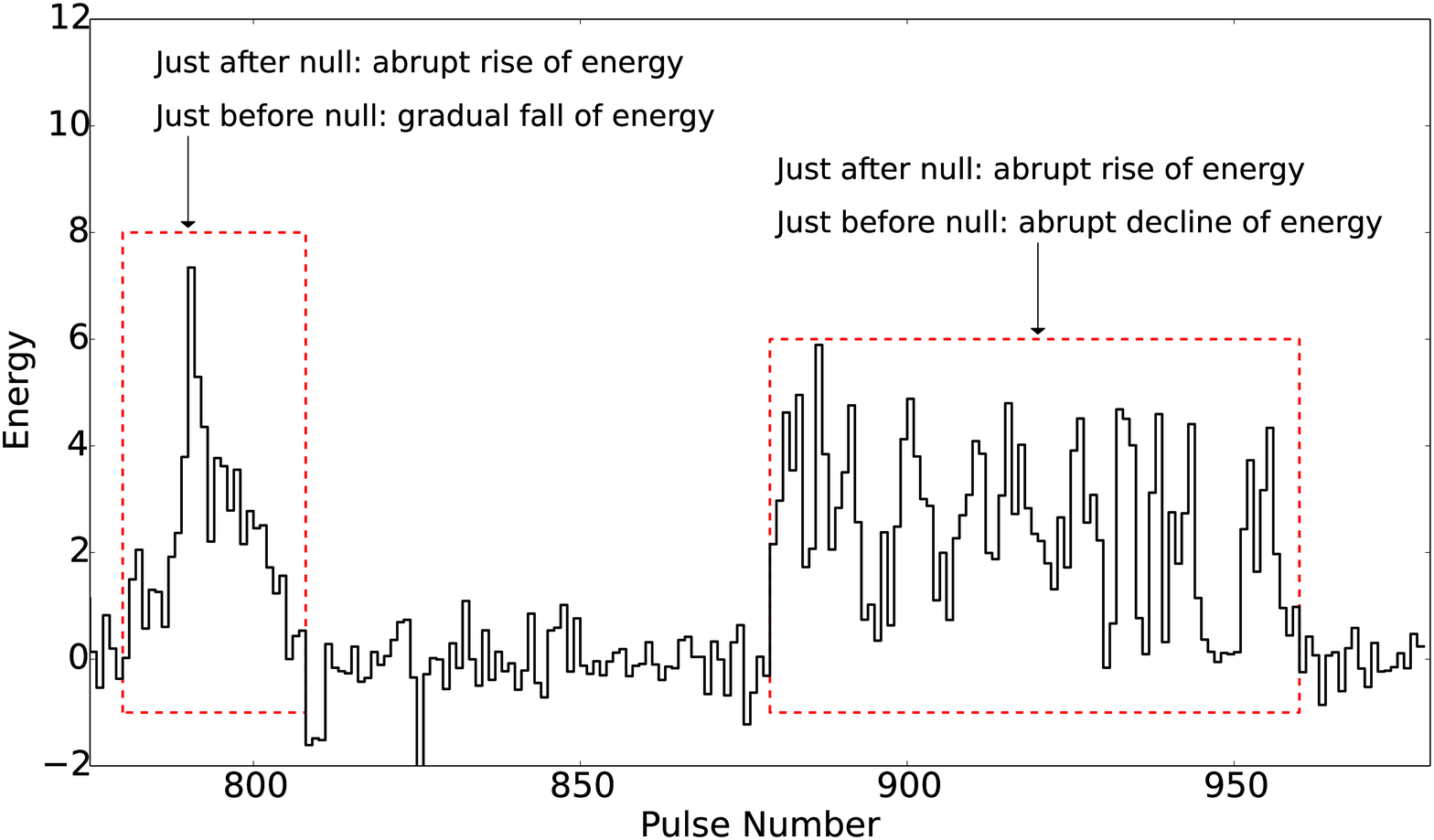}
\caption{Zoom-in view for the on-pulse energy versus pulse number from
  pulse 775 to 980 showing the two types of burst decay.}
\label{variation}
\end{center}
\end{figure}

Investigation of the transitional patterns between burst and null
states reveals different modes for the pulse energy changes. A zoom-in
view of pulse energy for the pulses from 775 to 980 is presented in
Fig. \ref{variation}, during which two burst states are detected. It
shows that the pulse intensity for the first few active pulses just
after nulls increases relatively rapidly. However, the transitions
from burst to null state shows two types of variation, either a
gradual decline as seen in the first burst, or a more abrupt
drop as seen in the second burst. These two transitional patterns
are reported here for the first time for this pulsar.

\begin{figure*}
\begin{center}
\includegraphics[width=80mm,height=77mm]{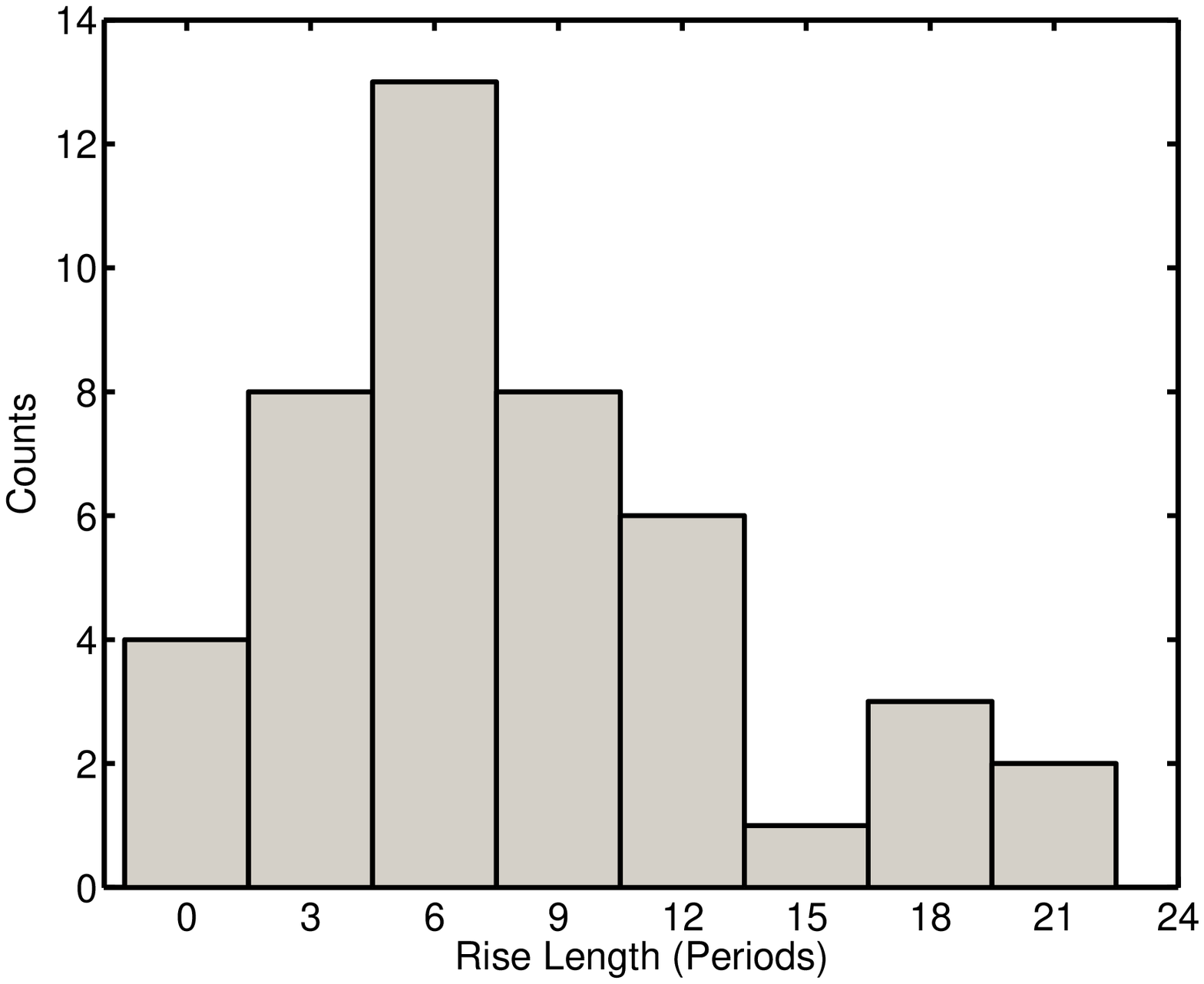}
\hspace{3ex}
\includegraphics[width=80mm,height=77mm]{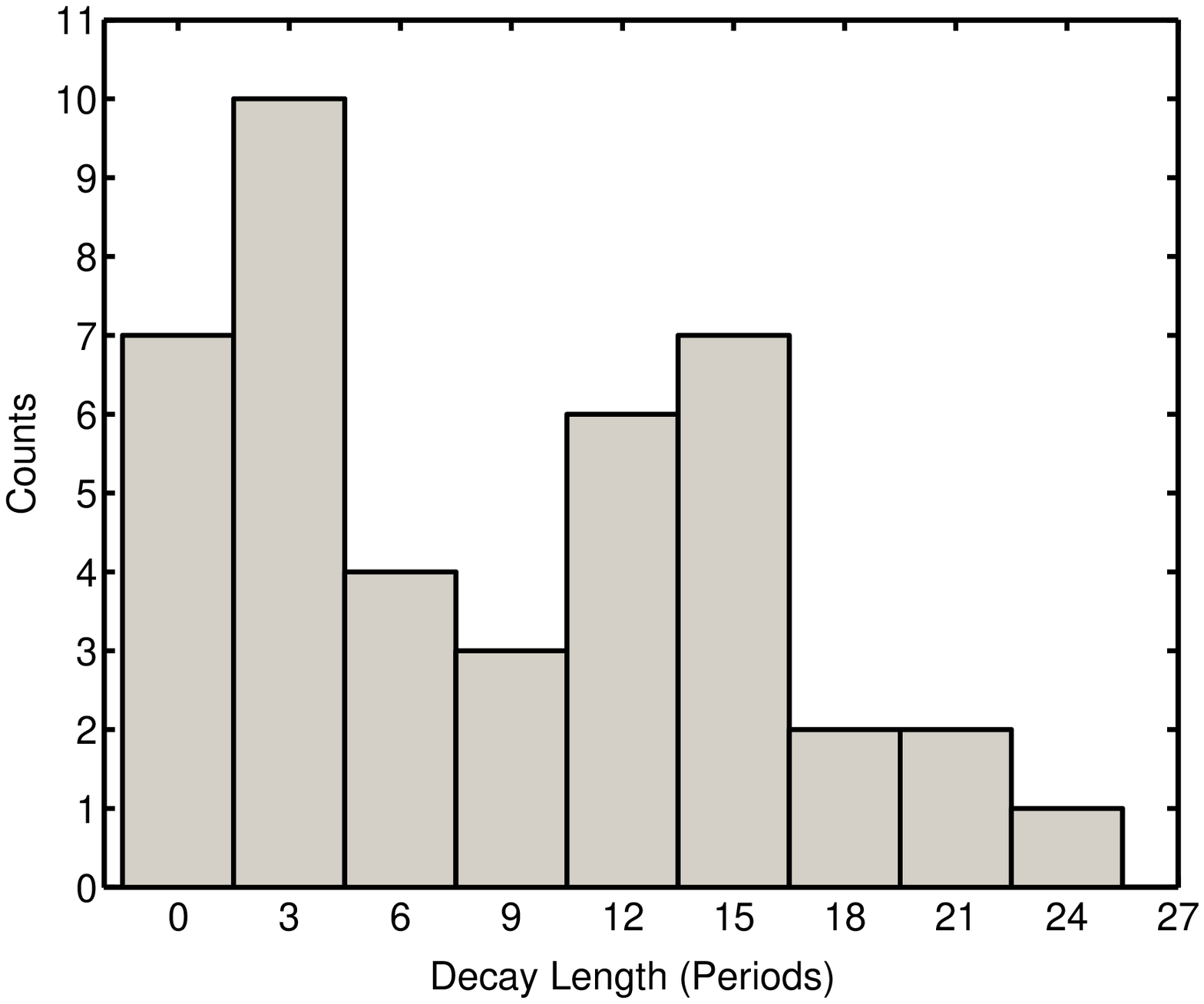}
\caption{Histograms of rise time (left panel) and fall time (right
  panel) from exponential fits to pulse energies at the start and end
  of bursts. }
\label{length}
\end{center}
\end{figure*}

The gradual rise at the onset of a burst and the slow decay
  from a burst to the null state are modeled by exponential functions
  \citep{lew04,vis14}. The rise and fall times are derived from a
  least-square fit of an exponential function to the data and are
  defined as the time between the start point and the point where the
  function is 90 per cent or 10 per cent of the maximum, for the rise
  and fall respectively. Fig. \ref{length} shows the resulting
histograms for 48 burst states with high S/N burst pulses. The rise
times are dominantly around six pulse periods, whereas a bi-modal
distribution is found for the decay times, with a prominent peak at
about three pulse periods and a second peak around 15 pulse periods.

\begin{figure}
\begin{center}
\includegraphics[width=95mm]{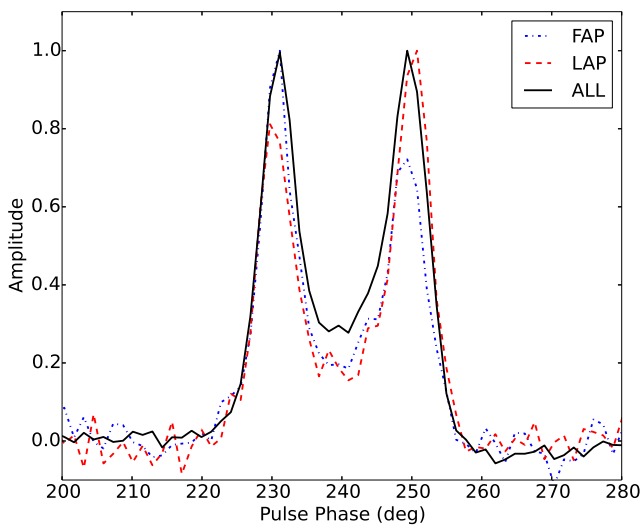}
\caption{The pulse profiles averaged from the first active pulse
  (dash-dotted line) and last active pulse (dashed line) of the burst
  states. The solid line represents the average pulse profile of all
  pulses. They are normalized by their respective peak intensities
  for comparison.}
\label{first_last}
\end{center}
\end{figure}

Fig. \ref{first_last} presents the average pulse profiles for the
first active pulse (FAP, shown with the dash-dotted line) immediately
after a null and the last active pulse (LAP, shown with the dashed
line) just before a null respectively. Both profiles have double
peaked components with similar pulse width, but the leading component
is stronger than the trailing component for FAP, and vice verse for
LAP. We applied the two-sample Kolmogorov-Smirnov (KS) test
  to the cumulative distribution functions of the FAP and LAP
  \citep{vis14} and found an 83 per cent probability that the two
  profiles were drawn from the same distribution.

\subsection{Identification of two subpulse drift states}

\begin{figure}
\begin{center}
\includegraphics[width=80mm,height=80mm]{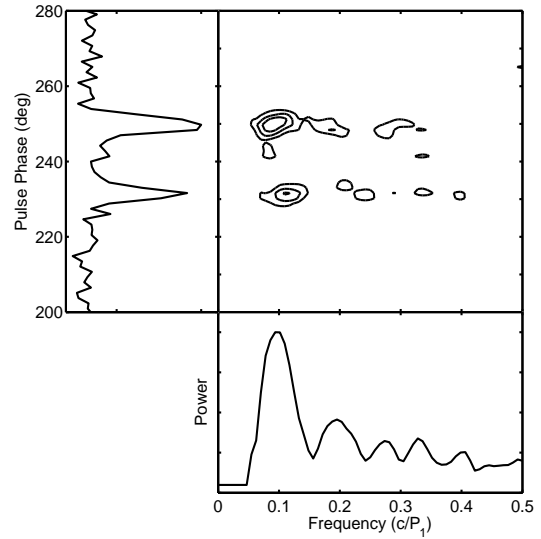}
\caption{Contour plot for the power spectrum of the flux density as a
  function of pulse phase for a sequence of 30 pulses showing a 0.1
  $c/P$ periodicity. The left-hand panel shows the power integrated
  over frequency. The bottom panel presents the power integrated over
  pulse phase. Signals at frequencies less than 0.05
  $c/P$ have been set to 0.}
\label{modeA}
\end{center}
\end{figure}

\begin{figure}
\begin{center}
\includegraphics[width=80mm,height=80mm]{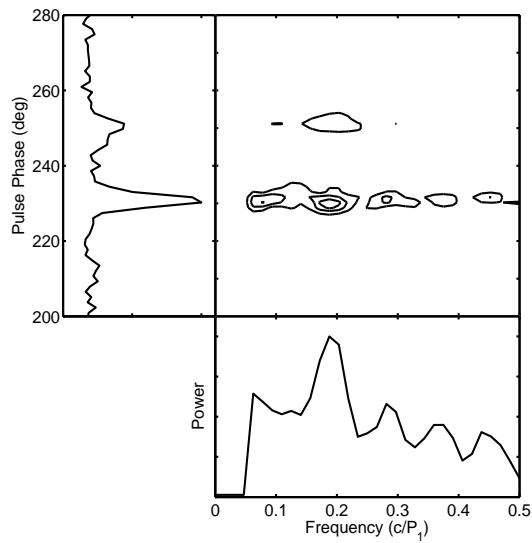}
\caption{Same as in Fig. \ref{modeA}, but for another pulse sequence
  of length 25 pulse periods where
  a 0.2 $c/P$ periodicity is detected prominently in the first
  component.}
\label{modeB}
\end{center}
\end{figure}

We used the phase-averaged power spectrum (hereafter PAPS) to analyse the
data for drifting subpulses. This method is suitable for
distinguishing drifting subpulses even when S/N is very low
\citep{smi05}. The contour plots shown in the main panels of
Fig. \ref{modeA} and \ref{modeB} are obtained from calculating the
absolute values of Fourier transform for flux density at each fixed
pulse phase across the pulse window. Subsequently, the PAPS is
obtained by averaging the resulted transforms over pulse phase. A
low-frequency modulation has been noted in numerous fluctuation studies
and therefore the signals with frequency less than 0.05 cycles per
rotational period (hereafter $c/P$) are set to 0 for clarity. In our
data reduction, the resulting PAPS is plotted from 0 up to 0.5 $c/P$
with a frequency resolution given by the reciprocal of the length of
the continuous pulse sequence. The width of 50 per cent of the peak is
taken as the uncertainty corresponding to a probability of larger than 68.27\%.

Figure \ref{modeA} shows an example of the power spectrum of the flux
as a function of pulse phase as well as the PAPS of a sequence of 30
pulses containing $10\ P$ periodicity. The PAPS peaks at
$0.10 \pm 0.02 \ c/P$. We classify this as mode A drift. The fluctuation
feature modulates all the emission in the leading and trailing region.

Figure \ref{modeB} shows the power spectrum for the flux density as a
function of pulse phase as well as the PAPS for a sequence of 25
single pulses, containing $5\ P$ periodicity. The PAPS peaks at
$0.21 \pm 0.04 \ c/P$. We classify this as mode B drift. It is noted that the
modulation feature is predominant in the leading component.

\begin{figure}
\begin{center}
\includegraphics[width=70mm,height=80mm]{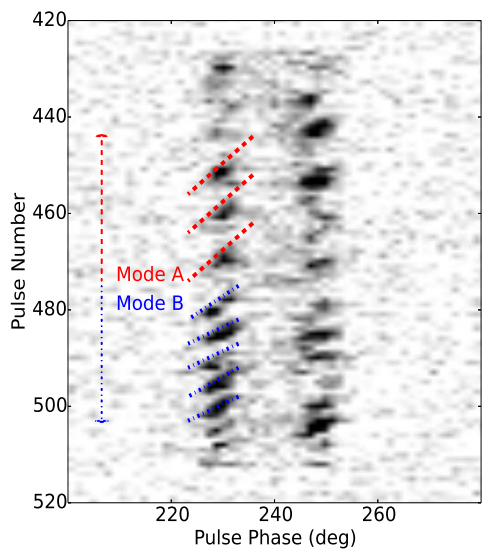}
\caption{Grey-scale plot of individual pulses showing an example of
  the rapid change in drift mode from A to B within a few pulses. The
  two drift modes are indicated by the dashed lines.}
\label{transform}
\end{center}
\end{figure}

An example of how fast the drift rate can change is shown in
Fig. \ref{transform}. As in Fig. \ref{grey_plot}, pulse number
increases from top to bottom. Mode A drifting is seen in the first
four drift bands, then within a few pulses, the drift switches to mode
B without any null. In addition, the first few active pulses just
after a null and the last few active pulses before a null show an
irregular drifting pattern, whereas central pulses have a more normal
drift pattern. This kind of transition in drift rates between burst
and null states is seen throughout most of the data set.

A careful examination of the subpulse drifting in the two components
reveals an irregular drifting in the trailing region (see pulses from
number 475 to 500, which exhibit subpulse drifting in mode B) on
several occasions, while the drifting in the leading region is
generally regular. The irregular drifting in the trailing component is
especially seen in Mode B and results in the weaker spectral feature
seen in Fig. \ref{modeB}.

\subsection{Comparison of two drift modes}

The periodicity of subpulse drifting is first checked visually
  from the single pulse sequence of each burst state. We then adjust
  the start and end of the sequence in order to achieve highest S/N of
  the PAPS peak. Finally, the location of the peak is determined as
  the drift frequency. Values of $\Delta \phi$ are determined by using
  the cross-correlation technique described by \citet{smi05} in which
  the correlation between consecutive pulses is fitted by a Gaussian
  curve. $P_2$ is then calculated by multiplying the mean phase drift
  by $P_3$.

\begin{figure*}
\begin{center}
\includegraphics[width=120mm]{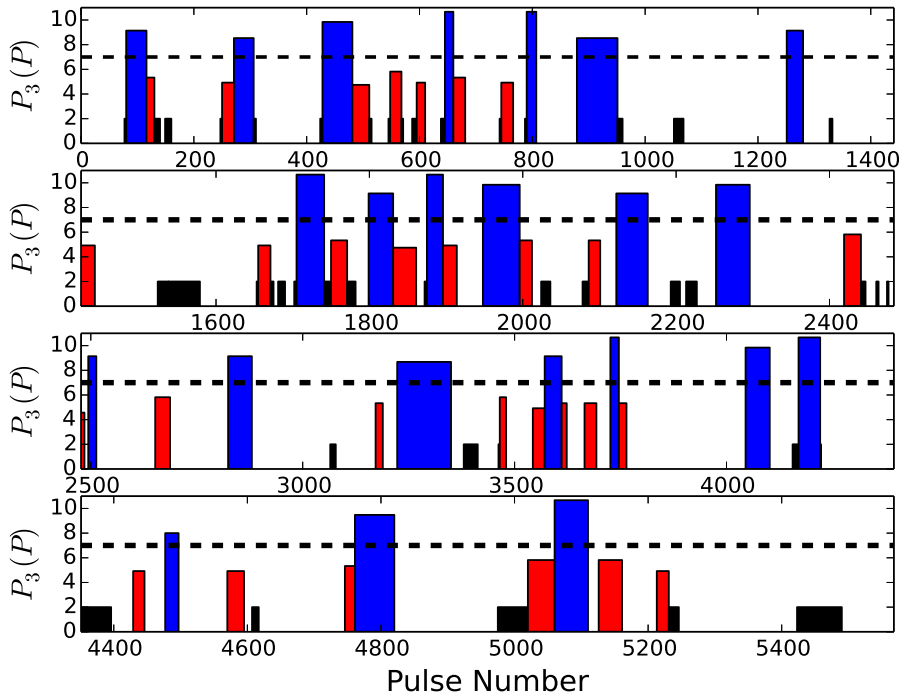}
\caption{Sequences of burst states giving observed values of
  $P_3$. Modes A and B are shown with blue and red bars, 
  and distinguished by the dashed lines. The black
  rectangles indicate bursts with no detection of subpulse
  drifting and the blank areas are nulls.}
\label{p3}
\end{center}
\end{figure*}

The values of $P_3$ for all drifting sequences in the 5568 pulses are shown in
Figure \ref{p3}. The black rectangles indicate sequence of burst
pulses without any detection of subpulse drifting, which we classify
as mode C for convenience. The blank regions represent nulls. The two
types of drifting subpulses are shown with blue and red bars. The transitions from one mode to the other
can be very rapid in one or a few pulse periods with no intervening
null. Table~1 summarises the occurence of the three modes
and the drift parameters for modes A and B. The measured $P_2$ for
drift mode A is larger than that for drift mode B by a factor of 1.43,
unlike PSR B0031$-$07 where $P_2$ for the three different drift modes
are the same \citep{hu70,smi05}.

\begin{table*}
\label{p3values}
 \centering
 \begin{minipage}{140mm}
  \caption{List of average values for $P_3$, $\Delta\phi$, $P_2$, and widths of the average intensity profiles for different drift modes at 1518 \rm{MHz} observed with Parkes radio telescope. The values of $P_2$ are derived from those of $P_3$ and $\Delta\phi$.}
  \begin{tabular}{@{}lccccccc@{}}
  \hline
   Drift & Number of& $P_3$& $\Delta\phi$& $P_2$& Number of& 10\% width& 50\% width\\
   mode  & sequences& ($P$)& ($\degr/P$)& ($\degr$)& pulses& ($\degr$)& ($\degr$)\\
 \hline
  A & 23 & $9.7\pm1.6$& $1.6\pm0.3$& $15.6\pm5.5$& 1019& $29.8\pm1.5$& $24.4\pm1.5$\\
  B &  29& $5.2\pm0.9$& $2.1\pm0.3$& $10.9\pm3.5$& 609& $30.7\pm1.5$& $25.0\pm1.5$\\
  C &  40& & & & 457& $31.5\pm1.5$& $24.6\pm1.5$\\
\hline
\end{tabular}
\end{minipage}
\end{table*}

\begin{figure}
\begin{center}
\includegraphics[width=80mm,height=60mm]{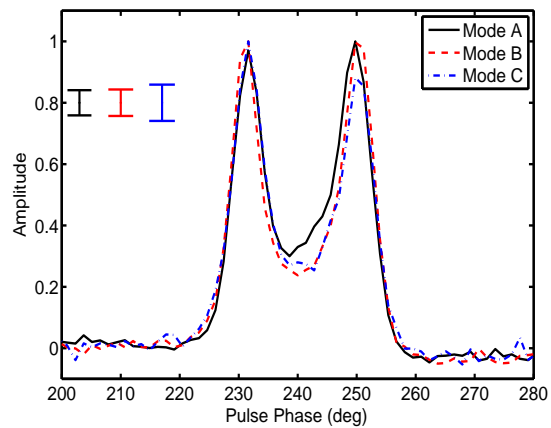}
\caption{The average pulse profiles for mode A (solid line), mode B
  (dashed line) and mode C (dash-dotted line), with $3\sigma$
  uncertainties determined from the off-pulse window.}
\label{modes}
\end{center}
\end{figure}

Figure \ref{modes} presents three average profiles obtained from
integrating pulses for each drift modes. The solid line shows the
average profile of 1019 pulses containing subpulses with mode A drift,
the dashed line shows the average profile of 609 pulses for mode B
drift, and 457 pulses for mode C drift is presented with the dash-dotted
line. The intensity of the leading and trailing components in drift
modes A and B are almost the same. For mode C, with no subpulse
drifting, the leading component is slightly stronger than the trailing
component in intensity. KS-test comparisons between the
average pulse profiles of the three modes suggest similar distributions
with probabilities of 89\% between A and C, 99\% between B and C, and
89\% between A and B. The last two columns in Table~1 give the
widths of the average intensity profiles for different modes. The profile
widths for the three drift modes are equal within the uncertainties. 

\begin{figure}
\begin{center}
\includegraphics[width=80mm,height=60mm]{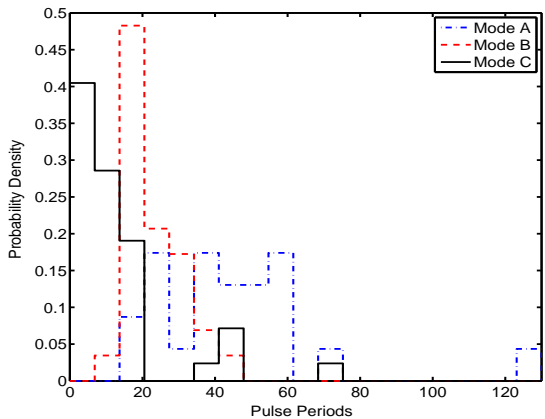}
\caption{Distributions of burst durations for the three modes. }
\label{density}
\end{center}
\end{figure}

Burst durations for each of the three different drift modes can be
obtained from examination of the single pulse data using the
above-mentioned method. Fig. \ref{density} shows these distributions
for the three drift modes, which can be compared with the overall
distribution of burst duration given by the upper plot in
Fig. \ref{duration}. On average, mode A bursts have longer durations
than mode B bursts, and mode C bursts have the shortest durations,
with about 90 per cent of mode C bursts having a duration of less than
20 pulse periods.

\section{Discussion and Conclusions}
\label{sec3}
We report new results on the properties of nulling and drifting
subpulses for pulsar J1727$-$2739 based on detailed analysis of 5568
pulse periods observed at centre frequency 1518~MHz using
the Parkes 64-m radio telescope. Nulls are detected, with durations of
6 to 281 pulse periods and an average null fraction of about
68\%. Burst durations range from 2 to 133 pulses, although occasional
isolated pulses are seen in otherwise null intervals. The intensity of
most bursts immediately after nulls is observed to rise to near
maximum in about six pulse periods, whereas the transition from bursts
to nulls exhibits two types of decay, either gradual or
rapid. Emission observed from the pulse profile averaged over all null
pulses reveals no pulsed emission with S/N greater than 3. Analysis on
the shape of the average profiles for last active pulses (LAP) and
first active pulses (FAP) reveal that they both have two components
and the same overall width, but LAP show a stronger trailing component
and opposite for FAP. This is inconsistent with previous studies which
suggested similarities between LAP and FAP \citep{viv95}. The observed
differences between LAP and FAP from J1727$-$2739 suggest that the
emission conditions are different at the start and end of
bursts. 

We demonstrate that PSR J1727$-$2739 exhibits drifting subpulses of
two distinct modes with characteristic ${P_3}$ of $9.7 \pm 1.6 P$ (mode A) and
$5.2 \pm 0.9 P$ (mode B), together with mode C in which no drifting subpulses
are detected. The occurrence rate was 49\% for mode A, 29\% for mode B
and 22\% for mode C as shown in Fig. \ref{density}. Our results also
show that transitions between different modes occur either rapidly
within a few pulses or are split by nulls. The onset and ending pulses
in the burst states show irregular drift properties, and occasional
irregular drifting patterns are observed in the trailing
component. Furthermore, the average pulse profiles for modes A and B
are similar with nearly equal components, but for mode C the leading
component is stronger.

Fig. \ref{p3} shows that some transitions between drift modes
  are rapid within a burst, while others transition after a null. This
  contrasts with PSRs B0809+74 and B0031$-$07 where changes in the
  drift pattern normally occur in association with a null. This
implies that for PSR J1727$-$2739 the nulls and the disordered drift
mode C do not act as so-called `reset phases' \citep{ran13}. However,
the drift rate reduces to zero near the onset and the end of most
burst states. Cross-correlation analyses between drift rate
  and average intensity and burst duration showed no significant
  correlations.

A modified model based on the traditional vacuum gap model
\citep{rs75} was proposed by \citet{gil03} with partial outflow of
thermal ions or electrons from the polar cap along with the
magnetospheric electron-positron pair plasma production. The
conductivity increases towards the end of the burst, with enough
charged particles in the accelerating gap to decrease the electric
field to a stable value where the radio emission is still active but
the drift rate goes off. With more accumulation, the acceleration gap
will not survive and thus the radio emission ceases.  The two patterns
of decay and the abrupt rise in pulse intensity before and after a
null also suggests different magnetospheric states in the
magnetosphere \citep{lyn10, mel14}. Switching between two different
magnetospheric states may result in the transitions between bursts and
null states, and the exponential gradual decay may be interpreted as
slow relaxation from one state to another. \citet{wan07} suggested
that emission can cease or commence suddenly when the charge or
magnetic configuration in the magnetosphere reaches a so-called
`tipping point' but the triggering mechanism is unknown.

The different observed subpulse drift modes suggest possible
variations in the emitting properties and may be an indication of a
local change in the electromagnetic field configuration in the
gap. Subpulses at different heights within same magnetic flux tube
will experience different plasma flow rates \citep{van03} resulting in
different drift modes. Another model by \citet{smi05} suggests that
groups of subpulses in different drift modes are located in different
concentric radiating rings and emitted from different magnetic field
lines. In this model, plasma at inner field lines relative to the
magnetic axis will experience greater electric potential giving faster
drift and stronger radio emission. Changes in drift mode result from
shifting of emission between different field lines. Thus both models
predict narrower average pulse profile for drift mode with emission
that comes from closer to the magnetic axis. However, the equal
observed pulse widths in the two drift modes implies that emission does
not shift relative to the magnetic axis. For magnetospheric plasma of
resistive nature \citep{2012ApJ...746...60L}, where the differential
rotation of plasma is responsible for the drifting emission in phase
and the conductivity determines the observed subpulse drifting
features, mode B may correspond to a faster ${\bf E} \times
{\bf B}$ drift and stronger emission because of
larger conductivity. However, as for the transitions between nulling
and burst states, the trigger mechanism for variations of conductivity
is still unclear.

Our results show that the time-scale for the change of the
magnetospheric current varies from several to dozens pulse periods,
and it is not yet clear as to how the emitted pulses are modulated in
intensity and in phase by the switching process of the magnetospheric
currents and the electric field. To the best of our knowledge, no
model can explain the observed features satisfactorily. It is
difficult to draw a definitive conclusion on the interaction between
nulling and switching of different subpulse drift modes. To continue
this study, we would require multi-frequency simultaneous single pulse
observations with full Stokes parameters to reveal more detailed
information on the variations of the size and location of the active
region.

\begin{acknowledgements}
We are grateful to the referee for valuable suggestions.
This work was supported by National Basic Research Program of China grants 973 Programs 2012CB821801 and 2015CB857100, the Pilot-B project grant XDB09010203, and the West Light Foundation of Chinese Academy of Sciences (WLFC) No.XBBS201422. We would like to thank members of the Pulsar Group at XAO for helpful discussion, and George Hobbs for useful comments and advice on the manuscript. JPY is funded by the National Natural Science Foundation of China (NSFC) under No.11173041. WMY is supported by NSFC under No.11203063, No.11273051, and the WLFC XBBS201123. VG acknowledge WLFC XBBS-2014-21. RY acknowledges supports from Project 11573059 NSFC. The Parkes radio telescope is part of the Australia Telescope which is funded by the Commonwealth Government for operation as a National Facility managed by the Commonwealth Scientific and Industrial Research Organization.
\end{acknowledgements}



\end{document}